\documentclass[american,aps,prl,reprint,superscriptaddress,twocolumn]{revtex4-1}
\usepackage[unicode=true,pdfusetitle, bookmarks=true,bookmarksnumbered=false,bookmarksopen=false, breaklinks=false,pdfborder={0 0 0},backref=false,colorlinks=false] {hyperref}
\hypersetup{ colorlinks,linkcolor=myurlcolor,citecolor=myurlcolor,urlcolor=myurlcolor}
\usepackage{braket,colortbl,cleveref,amsthm,amsmath,amssymb,txfonts}
\definecolor{myurlcolor}{rgb}{0,0,0.7}
\usepackage{graphics,graphicx}
\usepackage{color}

\theoremstyle{plain}

\def\bea{\begin{eqnarray}}
\def\eea{\end{eqnarray}}
\def\ba{\begin{array}}
\def\ea{\end{array}}

\def\ket{\rangle}
\def\bra{\langle}
\def\beq{\begin{equation}}
\def\eeq{\end{equation}}

\begin{document}

\title{ Convex geometry of Markovian Lindblad dynamics and witnessing non-Markovianity}

\author{Bihalan Bhattacharya}
\email{bihalan@gmail.com}
\author{Samyadeb Bhattacharya} 
\email{sbh.phys@gmail.com}
\affiliation{S. N. Bose National Centre for Basic Sciences, Block JD, Sector III, Salt Lake, Kolkata 700 098, India}

\begin{abstract}
\noindent  We develop a theory of linear witnesses for detecting non-Markovianity, based on the geometric structure of the set of Choi states for all Markovian evolutions having Lindblad type generators. We show that the set of all such Markovian Choi states form a convex and compact set under the small time interval approximation. Invoking geometric Hahn-Banach theorem, we construct linear witnesses to separate a given non-Markovian Choi state from the set of Markovian Choi states. We present examples of such witnesses for dephasing channel and Pauli channel in case of qubits. We further investigate the geometric structure of the Markovian Choi states to find that they do not form a polytope. This presents a platform to consider non-linear improvement of non-Markovianity witnesses.

\end{abstract}

\maketitle

\textit{Introduction:}
The theory of open quantum systems provides a powerful tool to study system-environment interactions, spawning decoherence, dissipation and other irreversible phenomena \citep{alicki,breuer}. In recent times,  much efforts have been devoted to characterize and classify quantum analogue of non-Markovian (NM) evolutions \citep{rivas1,breuerN,alonso,blp1,rhp1,bellomo,arend,samya1,samya2,samya3,wolf1,nm1,nm2,nm3,nm4,nm5,nm6}. It has been shown from various information theoretic and thermodynamic aspects, that NM can act as a powerful resource \citep{nmr,samya1,samya2,samya3}. However, it still remains a difficult yet important task to construct a theory of distinguishing NM evolutions from its Markovian counterparts, with proposals of  experimentally feasible detection procedures.  Very recently the present authors have constructed a convex resource theory of NM \citep{samya4}, creating that very opportunity of experimental verification, by exploring the geometry of NM dynamics in a similar manner of entanglement detection.

 The phenomena of witnessing entanglement \citep{wit1,wit2,wit3,wit4,wit5,wit6,wit7,wit8,wit9,wit10} is a stepping stone in the study of quantum information.  Witnesses are versatile tools for experimental detection of entangled quantum states. They are hermitian operators and hence observables by construction, giving positive expectation values for all separable states; whereas negativity of the same signifies the existence of entanglement. In this letter, we apply the tools of entanglement theory in open system dynamics, to construct NM witnesses from the similar footings of that of entanglement. 

Of late, works has been done to develop NM witnesses mimicking the same from entanglement theory \citep{NMwit,witnm1,witnm2}. But in order to construct a proper NM detection theory, we need to have a convex and compact set of states beholding the complete set of Markovian divisible operations. Though channel state duality \citep{jamil,choi} allows us to construct the set of states, due to the non-convex nature of divisible operations \citep{wolf1,wolf2}, constructing a theory of  linear witnesses is not possible in general.  We overcome this difficulty by ``small time interval" approximation, whence constructing the Choi states. This allows us to build a proper framework of linear witnesses for NM detection.

Before going into the main results of this work, we first elucidate the properties of the set of Choi states for divisible operations. Then we develop the theory of linear NM witnessing. We further consider the geometry of the set of Choi states, to identify the possibility of generalized non-linear witnesses for NM detection. Then we conclude with stating the possible implications. 

It is also very important here to mention that we are restricted to the set of operations having Lindblad type generators \citep{lindblad,gorini} of the form $\dot{\rho}(t)=\mathcal{L}(\rho(t))=\sum_{\alpha=1}^{n\leq d_S^2}\Gamma_{\alpha}(t)\left(L_{\alpha}\rho(t)L_{\alpha}^{\dagger}-\frac{1}{2}L_{\alpha}^{\dagger}L_{\alpha}\rho(t)-\frac{1}{2}\rho(t)L_{\alpha}^{\dagger}L_{\alpha}\right)$, with $\Gamma_{\alpha}(t)$s as the Lindblad coefficients, $A_{\alpha}$s as the Lindblad operators for a system with dimension $d_s$. For divisible operations we have $\Gamma_{\alpha}(t)\geq 0,~ \forall \alpha, t$ \citep{rhp1}.

\textit{Structure of Choi states:} Consider a general quantum channel $\Lambda(t,0):\rho(0)\rightarrow\rho(t)$. We construct a set $\mathbb{D}_{\mathcal{C}}$ including all such channels having Lindblad type generators. Exploiting channel-state duality, we define a one to one connection between $\mathbb{D}_{\mathcal{C}}$ and the set of corresponding Choi states $\mathbb{F}_{\mathcal{C}}$. A channel $\Lambda_{\mathcal{M}}(t,0)$ is CP-divisible iff $\Lambda_{\mathcal{M}}(t_3,t_1)=\Lambda_{\mathcal{M}}(t_3,t_2)\circ\Lambda_{\mathcal{M}}(t_2,t_1)$, with $t_3>t_2>t_1$ $\forall t_1, t_2, t_3$. It implies that the Choi state $\mathcal{C}_{\mathcal{M}}(t_2,t_1)=\mathbb{I}\otimes\Lambda(t_2,t_1)|\phi\ket\bra\phi|$ is a valid density matrix for every instant of time, with $||\mathcal{C}_{\mathcal{M}}(t_2,t_1)||_1=1$, $\forall t_1,t_2$. Here $|\phi\ket$ corresponds to a maximally entangled state in $d_s^2$ dimension and $||\cdot||_1=Tr[\sqrt{(\cdot)^{\dagger}(\cdot)}]$ is the trace norm. CP-divisibility breaking of a channel signifies NM backflow of information \citep{rhp1,blp1}. The set of all CP-divisible channels $\mathbb{D}_{\mathcal{M}}\subset\mathbb{D}_{\mathcal{C}} $ is then considered to be the set of all Markovian memoryless channels. We therefore define a subset $\mathbb{F}_{\mathcal{M}}\subset\mathbb{F}_{\mathcal{C}}$ as $\mathbb{F}_{\mathcal{M}}=\{\mathcal{C}_{\mathcal{M}}(t_2, t_1)~|~~ ||\mathcal{C}_{\mathcal{M}}(t_2, t_1)||_1=1, \forall t_1, t_2\}$, including the Choi states for all CP-divisible operations and prove the following proposition. 

\textbf{Proposition 1:} The set of Markovian Choi states (MCS)  $\mathbb{F}_{\mathcal{M}}^{\epsilon}=\{\mathcal{C}_{\mathcal{M}}(t+\epsilon, t)~|~~ ||\mathcal{C}_{\mathcal{M}}(t+\epsilon, t)||_1=1, \forall t, \epsilon\}$ is a convex and compact set in the limit $\epsilon\rightarrow 0$.

\textit{Proof:} Let $\mathcal{C}_{\mathcal{M}}^{(1)}(t+\epsilon, t)$ and  $\mathcal{C}_{\mathcal{M}}^{(2)}(t+\epsilon, t)$ be two MCS  corresponding to two separate Markovian operation $\Lambda_{\mathcal{M}}^{(1)}$ and $\Lambda_{\mathcal{M}}^{(2)}$ having Lindblad type generator $\mathcal{L}_t^{(1)}$ and $\mathcal{L}_t^{(2)}$
 with positive coefficient respectively. Therefore we have,
$
\Lambda_{\mathcal{M}}^{(i)}(t+\epsilon,t)\equiv \exp\left(\int_{t}^{t+\epsilon}\mathcal{L}_t^{(i)}dt\right),
~\mbox{with}~i=1,2.
$
For sufficiently small $\epsilon$, expanding the exponential and neglecting 2nd order onward terms we have
$\Lambda_{\mathcal{M}}^{(i)}(t+\epsilon,t)=\mathbb{I}+\epsilon\mathcal{L}_t^{(i)},
~\mbox{with}~i=1,2.
$
Defining another map
$
\Lambda_{\mathcal{M}}(t+\epsilon,t)=p\Lambda_{\mathcal{M}}^{(1)}(t+\epsilon,t)+(1-p)\Lambda_{\mathcal{M}}^{(2)}(t+\epsilon,t)
=\mathbb{I}+\epsilon[p\mathcal{L}_t^{(1)}+(1-p)\mathcal{L}_t^{(2)}]=\mathbb{I}+\epsilon\mathcal{L}_t,
$
with $\mathcal{L}_t=p\mathcal{L}_t^{(1)}+(1-p)\mathcal{L}_t^{(2)}$ and $0\leq p\leq 1$, we get that $\mathcal{L}_t$ is also a Lindblad type generator with positive coefficients. This shows that the map $\Lambda_{\mathcal{M}}(t+\epsilon,t)$ also belongs to the set of divisible Markovian maps, proving $\mathbb{F}_{\mathcal{M}}^{\epsilon}$ is a convex set.

To prove $\mathbb{F}_{\mathcal{M}}^{\epsilon}$ is compact it is enough to show $\mathbb{F}_{\mathcal{M}}^{\epsilon}$ is closed and bounded in the concerned topology. This fact is obvious from the definition of $\mathbb{F}_{\mathcal{M}}^{\epsilon}$ and continuity of trace norm. \qed \\

We consider finite dimensional normed linear spaces only, where all norms are topologically equivalent. In light of this fundamental fact, evidently  
$\mathbb{F}_{\mathcal{M}}^{\epsilon}$ is also compact under Hilbert-Schmidt (HS) norm $||\cdot||_2=\sqrt{Tr[(\cdot)^{\dagger}(\cdot)]}$. Moreover, it can be shown that under the assumption of small time interval ($\epsilon\rightarrow 0$), the HS norm of any arbitrary Choi state is approximately 1. The argument follows like this.
 $||\mathcal{C}_{\mathcal{N}}||_2 \approx\sqrt{Tr[(|\phi\ket\bra\phi|+\epsilon\mathbb{I}\otimes\mathcal{L}(\phi\ket\bra\phi|))^{\dagger}(|\phi\ket\bra\phi|+\epsilon\mathbb{I}\otimes\mathcal{L}(\phi\ket\bra\phi|))]}$. Neglecting higher order terms of $\epsilon$, we have $||\mathcal{C}_{\mathcal{N}}||_2 \approx\sqrt{Tr[||\phi\ket\bra\phi|+\epsilon\mathbb{I}\otimes\mathcal{L}|\phi\ket\bra\phi|]}=1$, since $\mathcal{L}(\cdot)$ is always traceless.  Thus the structure of $\mathbb{F}_{\mathcal{M}}^{\epsilon}$ leads us to the existence of non-Markovianity witnesses, by exploiting techniques of convex analysis \cite{convex}.

It is an interesting question, that whether in principle a non-Markovian Choi state (NMCS) can be separated from all MCS. Invoking geometric Hahn-Banach separation theorem \citep{convex}, we show that the answer is affirmative.

\textbf{Theorem 1:} A NMCS can be separated from all MCS by a hyperplane.\\
\textit{Proof:} Let $\mathcal{C}_{\mathcal{N}}$ be a NMCS and $\mathbb{F}_{\mathcal{M}}^\epsilon$ denotes the set of all MCS. Note that,
$D(\mathcal{C}_{\mathcal{N}}$ $\vert$ $\mathbb{F}_{\mathcal{M}}^\epsilon )=0$  iff $\mathcal{C}_{\mathcal{N}}\in Cl(\mathbb{F}_{\mathcal{M}}^\epsilon)$, where $Cl(\cdot)$ denotes the topological closure. Since $\mathbb{F}_{\mathcal{M}}^\epsilon$ is closed as $\mathbb{F}_{\mathcal{M}}^\epsilon$ is compact, then Cl$(\mathbb{F}_{\mathcal{M}}^\epsilon) = \mathbb{F}_{\mathcal{M}}^\epsilon$. $\mathcal{C}_{\mathcal{N}}$ does not belong to $\mathbb{F}_{\mathcal{M}}^\epsilon$, thus $D(\mathcal{C}_{\mathcal{N}}$ $\vert$ $\mathbb{F}_{\mathcal{M}}^\epsilon )$ $>$ 0. Considering singleton set $\lbrace \mathcal{C}_{\mathcal{N}} \rbrace $ as convex set, we always have  a hyperplane \citep{convex} separating $\mathcal{C}_{\mathcal{N}}$ and $\mathbb{F}_{\mathcal{M}}^\epsilon$. \qed

\begin{figure}[htb]
	{\centerline{\includegraphics[width=8cm, height=6cm] {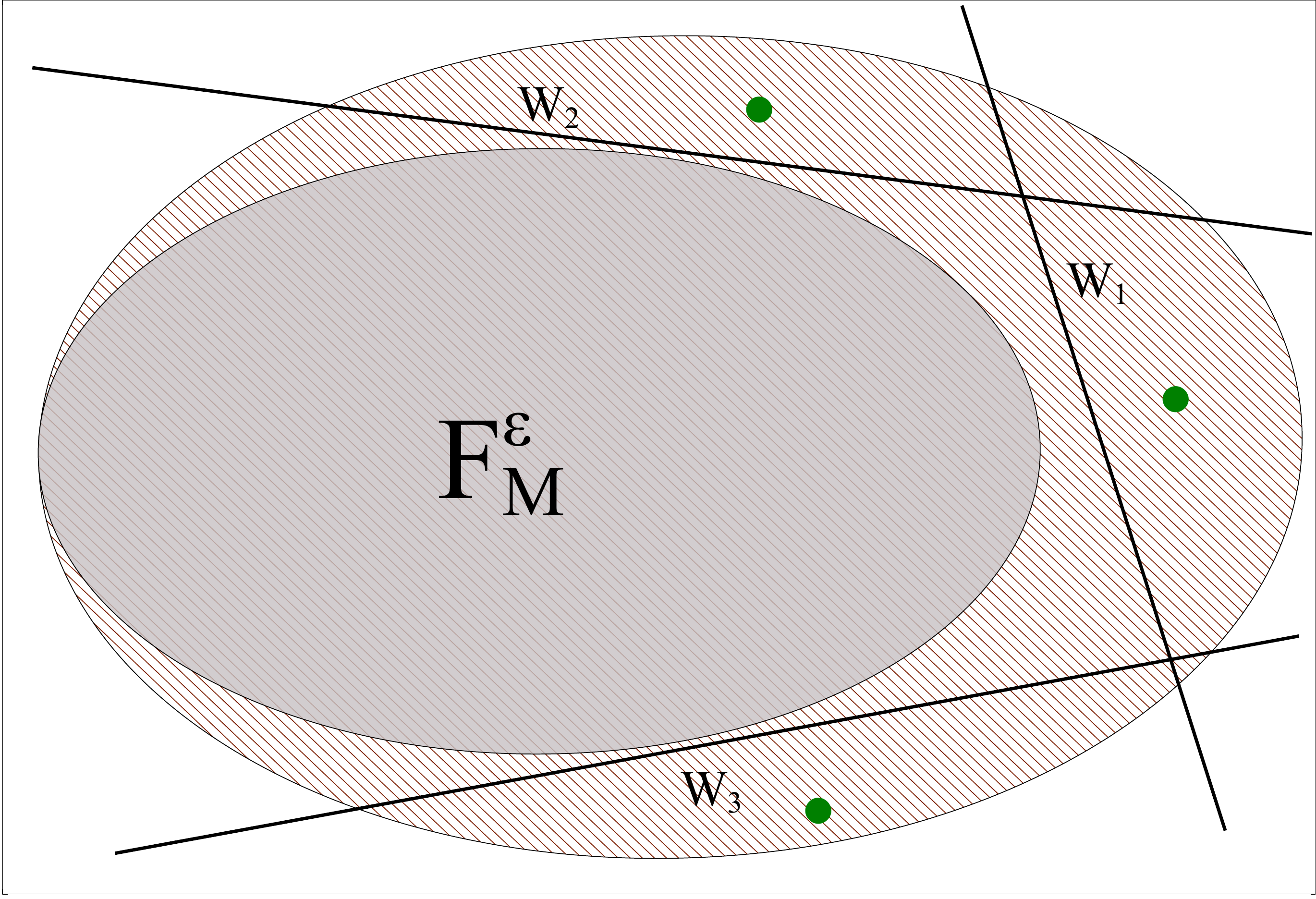}}}
	\caption{(Colour online) \\ Here the larger ellipse represents set of all Choi states and $\mathbb{F}_{\mathcal{M}}^\epsilon$ represents the convex compact set of MCS at sufficiently small time $\epsilon$ and $W_1, W_2,W_3$ represents three hyperplanes  separating three NMCS (represented as green dot) from MCS. }
	\label{fig2}

\end{figure}
In Fig. \eqref{fig2} we represent the schematic diagram for \textbf{Theorem 1}. Since $\mathbb{F}_{\mathcal{M}}^\epsilon$ is convex and compact, every NMCS can in principle be separated from $\mathbb{F}_{\mathcal{M}}^\epsilon$ by some separating hyperplane. 

\textit{Non-Markovianity Witness:}  Let us consider the construction of NM witness using the techniques of entanglement theory. 

\textit{Definition:} A hermitian operator $W$ is said to be a NM witness if it satisfies following criteria: 
\begin{enumerate}
\item[1.] $Tr(W \mathcal{C}_{\mathcal{M}} )$ $\geqslant$ 0 $\forall$ $\mathcal{C}_{\mathcal{M}}$ $\in$
$\mathbb{F}_{\mathcal{M}}^\epsilon$.\\
\item[2.] There exists atleast one NM Choi state $\mathcal{C}_{\mathcal{N}}$ such that $Tr(W \mathcal{C}_{\mathcal{N}} )$ $<$ 0.
\end{enumerate}

It is clear from the definition, that a single witness can not detect all NMCS. The witness will depend on the NMCS, which one wishes to detect.

\textit{Construction of NM witness:}
Let $\mathcal{C}_{\mathcal{Q}}$ be a finite dimensional Choi state corresponding to some operation $\mathcal{Q}$.    $\mathcal{C}_{\mathcal{Q}}$ being hermitian, we have its spectral decomposition as:
\begin{equation}\label{1}
\mathcal{C}_{\mathcal{Q}}=\sum \lambda_i P_i ,
\end{equation}
where $P_i$s are orthogonal projections onto the subspace spanned by the normalised eigenvectors corresponding to the eigenvalues $\lambda_i$. Note that,
$
Tr(\mathcal{C}_{\mathcal{Q}} P_j)= \lambda_j \delta_{ij},
$
with $\delta_{ij}$  being the Kronecker delta. If the operation is CP-divisible, then $\mathcal{C}_{\mathcal{Q}}$ being a valid state has all non negative eigenvalues. Hence, $Tr(\mathcal{C}_{\mathcal{Q}} P_j)$ $\geqslant$ 0 $\forall$ j. If the operation is NM, $Tr(\mathcal{C}_{\mathcal{Q}} P_j)$ $<$ 0 for atleast one $j$, as $\mathcal{C}_{\mathcal{Q}}$ has atleast one negative eigenvalue. Thus the orthogonal projectors serve as witnesses for  NMCS.

\textit{Examples of qubit channels:} Let us have two examples of NM witness based on our construction. \\
\textbf{Dephasing channel:} The Lindblad equation corresponding to a pure dephasing channel  is given by:
 $\dfrac{d\rho}{dt}$ = $\gamma(t)(\sigma_z\rho\sigma_z - \rho)$. Performing small time approximation, we get the eigenvalues of the corresponding Choi state as 0, 0, $\gamma(t) \epsilon$ and $1-\gamma(t)\epsilon$ respectively. Since $\epsilon$ is infinitesimally small, the later of the two non-negative eigenvalues can always be considered as positive. We therefore take the orthogonal projector corresponding to the eigenvalue $\gamma(t) \epsilon$ as the witness for NM since $\gamma(t)\geq 0$ for CP-divisible operations and can be negative in case of NM operations. The witness for qubit dephasing operation is therefore $P_{deph}=|\chi\ket\bra\chi |$, where $|\chi\ket=(0,~1,~0,~0)^T$.\\
\textbf{Pauli Channel:} Consider now the Pauli channel: $\dfrac{d\rho}{dt}$ = $\gamma_x(t)(\sigma_x\rho\sigma_x - \rho)$+$\gamma_y(t)(\sigma_y\rho\sigma_y - \rho)$+$\gamma_z(t)(\sigma_z\rho\sigma_z - \rho)$. Small time approximation gives the eigenvalues of the corresponding Choi state as $1-(\gamma_x(t)+\gamma_y(t)+\gamma_z(t))\epsilon$, $\gamma_x(t)\epsilon$, $\gamma_y(t)\epsilon$, $\gamma_z(t)\epsilon$. Following same logic,we can take the the orthogonal projectors corresponding to the eigenvalues $\gamma_i(t)\epsilon$ ( with $i=x,y,z$) as the witnesses of NM. They are respectively given by $P_x=|\chi_x\ket\bra\chi_x|$, $P_y=|\chi_y\ket\bra\chi_y|$ and $P_z=|\chi_z\ket\bra\chi_z|$, with $|\chi_x\ket=(0,~1,~1,~0)^T$,  $|\chi_y\ket=(0,-1,~1,~0)^T$ and  $|\chi_z\ket=(-1,~0,~0,~1)^T$.

\textit{Alternative construction of NM witness:} In the above mentioned procedure, the construction of witness depends on the eigenvalues of the corresponding Choi state. For a large dimensional system, computing the eigenvalues and the corresponding projectors may be difficult. Therefore we adopt an alternative formalism \citep{rubin} based on the structure of $\mathbb{F}_{\mathcal{M}}^\epsilon$.

In order to construct the witness, we first need to show the existence of nearest MCS corresponding to some NMCS. Consider $\mathcal{N}$  to be a NM operation having Choi state $\mathcal{C_{\mathcal{N}}}$. The distance between $\mathcal{C_{\mathcal{N}}}$ and $\mathbb{F}_{\mathcal{M}}^\epsilon$ is given by $M=inf_{\mathcal{C_{\mathcal{M}}} \in \mathbb{F}_{\mathcal{M}}^\epsilon } D(\mathcal{C_{\mathcal{N}}}\mid \mathcal{C_{\mathcal{M}}})$, where $D(\cdot\mid\cdot)$ is any proper metric. We now prove the following theorem.

\textbf{Theorem 2:} Corresponding to any NMCS, there always exists a nearest MCS.

\textit{Proof:} Fixing a NMCS $\mathcal{C_{\mathcal{N}}}$, we define a function $g: \mathbb{F}_{\mathcal{M}}^\epsilon$ $\rightarrow \mathbb{R}~$  by setting,
$g(\mathcal{C_{\mathcal{M}}})$=$D(\mathcal{C_{\mathcal{N}}} \mid \mathcal{C_{\mathcal{M}}})$     $\forall$  $\mathcal{C_{\mathcal{M}}}$ $\in$ $\mathbb{F}_{\mathcal{M}}^\epsilon$.
Clearly $g$ is a continuous function on the set $\mathbb{F}_{\mathcal{M}}^\epsilon$. Moreover since $\mathbb{F}_{\mathcal{M}}^\epsilon$ is compact, $\exists$ $\mathcal{C_{\mathcal{M_*}}}$  $\in$  $\mathbb{F}_{\mathcal{M}}^\epsilon$ such that $g(\mathcal{C_{\mathcal{M_*}}})$=$inf_{\mathcal{C_{\mathcal{M}}} \in \mathbb{F}_{\mathcal{M}}^\epsilon}$  $g({\mathcal{C^{\mathcal{M}}}})$. Hence infimum is achieved by some MCS.\qed 

Having proved the existence of nearest MCS, now it will be interesting to ask whether or under which condition the nearest MCS corresponding to a NMCS is unique. Thus follows the next proposition.

\textbf{Proposition 2:} Let $\mathcal{C_{\mathcal{N}}}$ be a NMCS. Then $\mathcal{C_{\mathcal{M^*}}}$ is the unique nearest MCS if and only if for all $\mathcal{C_{\mathcal{M}}}$ $\in$ $\mathbb{F}_{\mathcal{M}}^\epsilon$, $Tr[(\mathcal{C_{\mathcal{N}}}-\mathcal{C_{\mathcal{M^*}}})(\mathcal{C_{\mathcal{M}}}-\mathcal{C_{\mathcal{M^*}}})]$ $\leq$ 0. 

\textit{Proof:} To prove the sufficient part, let for all MCS $\mathcal{C_{\mathcal{M}}}$, $Tr[(\mathcal{C_{\mathcal{N}}}-\mathcal{C_{\mathcal{M^*}}})(\mathcal{C_{\mathcal{M}}}-\mathcal{C_{\mathcal{M^*}}})]$ $\leq$ 0. Considering $Tr[(\mathcal{C_{\mathcal{N}}}-\mathcal{C_{\mathcal{M}}})^{2}]$, and by adding and subtracting $\mathcal{C_{\mathcal{M^*}}}$  we get,
$
Tr[(\mathcal{C_{\mathcal{N}}}-\mathcal{C_{\mathcal{M}}})^{2}] -Tr[ (\mathcal{C_{\mathcal{N}}}-\mathcal{C_{\mathcal{M^*}}})^{2}] 
\geq -2 Tr[(\mathcal{C_{\mathcal{N}}}-\mathcal{C_{\mathcal{M^*}}})(\mathcal{C_{\mathcal{M}}}-\mathcal{C_{\mathcal{M^*}}})].
$
 Using the hypothesis we have $Tr[(\mathcal{C_{\mathcal{N}}}-\mathcal{C_{\mathcal{M}}})^{2}] \geq$ $Tr[(\mathcal{C_{\mathcal{N}}}-\mathcal{C_{\mathcal{M^*}}})^{2}]$. This shows $\mathcal{C_{\mathcal{M^*}}}$ is the nearest MCS corresponding to the NMCS $\mathcal{C_{\mathcal{N}}}$.

To prove the necessary part, let $\mathcal{C_{\mathcal{M^*}}}$ be the nearest MCS corresponding to NMCS $\mathcal{C_{\mathcal{N}}}$. Then,
$Tr[(\mathcal{C_{\mathcal{N}}}-\mathcal{C_{\mathcal{M}}})^{2}]$ $\geq$ $Tr[(\mathcal{C_{\mathcal{N}}}-\mathcal{C_{\mathcal{M^*}}})^{2}]$ for any MCS $\mathcal{C_{\mathcal{M}}}$ $\in$
$\mathbb{F}_{\mathcal{M}}^\epsilon$. This implies $Tr[(\mathcal{C_{\mathcal{N}}}-\mathcal{C_{\mathcal{M^*}}})(\mathcal{C_{\mathcal{M}}}-\mathcal{C_{\mathcal{M^*}}})]$ $\leq$ $\dfrac{1}{2}$ $Tr[(\mathcal{C_{\mathcal{M^*}}}-\mathcal{C_{\mathcal{M}}})^{2}]$. Since $\mathbb{F}_{\mathcal{M}}^\epsilon$ is convex, let $\mathcal{C_{\mathcal{M}}}$ = $(1-\mu)\mathcal{C_{\mathcal{Q}}}$ + $\mu \mathcal{C_{\mathcal{M^*}}}$ with 0 $<$ $\lambda$ $<$ 1, where $\mathcal{C_{\mathcal{Q}}}$ $\in$ $\mathbb{F}_{\mathcal{M}}^\epsilon$. Therefore  $Tr[(\mathcal{C_{\mathcal{N}}}-\mathcal{C_{\mathcal{M^*}}})((1-\mu)\mathcal{C_{\mathcal{Q}}} + \mu \mathcal{C_{\mathcal{M^*}}}-\mathcal{C_{\mathcal{M^*}}})]$ $\leq$ $\dfrac{1}{2}$ $Tr[((1-\mu)\mathcal{C_{\mathcal{Q}}}$ + $\mu \mathcal{C_{\mathcal{M^*}}}-\mathcal{C_{\mathcal{M^*}}})^{2}]$, which gives the inequality $Tr[(\mathcal{C_{\mathcal{N}}}-\mathcal{C_{\mathcal{M^*}}})(\mathcal{C_{\mathcal{Q}}}-\mathcal{C_{\mathcal{M^*}}})]$ $\leqslant$ $\dfrac{1}{2}(1-\mu)Tr[(\mathcal{C_{\mathcal{M^*}}}-\mathcal{C_{\mathcal{Q}}})^{2}]$. Letting $\mu$ $\rightarrow$ 1, we have the result.

 To prove the uniqueness of $\mathcal{C_{\mathcal{M^*}}}$, let $\mathcal{C_{\mathcal{M^1}}}$ be another MCS which minimizes $Tr[(\mathcal{C_{\mathcal{N}}}-\mathcal{C_{\mathcal{M}}})^{2}]$. Then $Tr[(\mathcal{C_{\mathcal{N}}}-\mathcal{C_{\mathcal{M^*}}})(\mathcal{C_{\mathcal{M^1}}}-\mathcal{C_{\mathcal{M^*}}})]$ $\leqslant$ 0 and $Tr[(\mathcal{C_{\mathcal{N}}}-\mathcal{C_{\mathcal{M^1}}})(\mathcal{C_{\mathcal{M^*}}}-\mathcal{C_{\mathcal{M^1}}})]$ $\leqslant$ 0 together implies $\mathcal{C_{\mathcal{M^*}}}$ = $\mathcal{C_{\mathcal{M^1}}}$, proving the uniqueness of the nearest MCS. \qed

\noindent Armed with this proposition, we now prove the following theorem.
 
\textbf{Theorem 3:}
Let $\mathcal{C_{\mathcal{M^*}}}$ be the nearest MCS to a NMCS $\mathcal{C_{\mathcal{N}}}$. Then
\begin{equation}\label{2}
\mathcal{W}=c_0\mathbb{I}+\mathcal{C_{\mathcal{M^*}}}-\mathcal{C_{\mathcal{N}}},
\end{equation}
with $c_0=Tr(\mathcal{C_{\mathcal{M^*}}}(\mathcal{C_{\mathcal{N}}}-\mathcal{C_{\mathcal{M^*}}}))$ is a NM witness for $\mathcal{C_{\mathcal{N}}}$. \\

\textit{Proof:} We verify that, $Tr[\mathcal{W}\mathcal{C_{\mathcal{S}}}] = -Tr[(\mathcal{C_{\mathcal{S}}}-\mathcal{C_{\mathcal{M^*}}})(\mathcal{C_{\mathcal{N}}}-\mathcal{C_{\mathcal{M^*}}})]$ for any Choi state $\mathcal{C_{\mathcal{S}}}$. To prove $\mathcal{W}$ is a NM witness for $\mathcal{C_{\mathcal{N}}}$ it is enough to show 1) $\mathcal{W}$ is hermitian, 2) $Tr[\mathcal{W}\mathcal{C_{\mathcal{M}}}]$ $\geqslant$ 0 $\forall$ $\mathcal{C_{\mathcal{M}}}$ $\in$ $\mathbb{F}_{\mathcal{M}}^\epsilon$ and $Tr[\mathcal{W}\mathcal{C_{\mathcal{N}}}]$ $<$ 0.\\
 $\mathcal{W}$ is hermitian according to its definition. Since $Tr[\mathcal{W}\mathcal{C_{\mathcal{S}}}] = -Tr[(\mathcal{C_{\mathcal{S}}}$-$\mathcal{C_{\mathcal{M^*}}})(\mathcal{C_{\mathcal{N}}}-\mathcal{C_{\mathcal{M^*}}})]$ holds for any Choi state, it also holds for any $\mathcal{C_{\mathcal{M}}}$ in $\mathbb{F}_{\mathcal{M}}^\epsilon$. 
 
Since $\mathcal{C_{\mathcal{M^*}}}$ is the nearest MCS for $\mathcal{C_{\mathcal{N}}}$, we have $Tr[(\mathcal{C_{\mathcal{N}}}-\mathcal{C_{\mathcal{M^*}}})(\mathcal{C_{\mathcal{M}}}-\mathcal{C_{\mathcal{M^*}}})]$ $\leqslant$ 0. Hence it follows that 
\[Tr[\mathcal{W}\mathcal{C_{\mathcal{M}}}] \geqslant 0~~ \forall~~ \mathcal{C_{\mathcal{M}}}\in\mathbb{F}_{\mathcal{M}}^\epsilon.\] 
Using the trace preservation property of quantum operations, we get \[ Tr[\mathcal{W}\mathcal{C_{\mathcal{N}}}]=-Tr(\mathcal{C_{\mathcal{N}}}-\mathcal{C_{\mathcal{M^*}}})^{2} < 0 \]  \qed
 
Having constructed the theory of linear witnesses to detect NM dynamics, we ask the immediate  following question that, whether linear witnesses are sufficient to determine all the NMCS. In the theory of entanglement detection, we know that non-linear improvement of witnesses gives us further advantages to detect entanglement \citep{nwit1,nwit2,nwit3}. In the following, we discuss such possibilities for NM detection.
 
 \textit{Possibility of non-linear witness:}  After constructing the structure of linear witnesses to detect the NMCS, a very legitimate question should be the following. How many linear  witnesses are enough to capture all NMCS. To answer this question, we need to investigate the geometry of $\mathbb{F}_{\mathcal{M}}^\epsilon$;  i.e. precisely whether the set $\mathbb{F}_{\mathcal{M}}^\epsilon$ forms a polytope determined by intersection of finitely many half-spaces  obtained from linear witnesses. In analogy to Ref. \citep{polytope}, we surmise that these finitely many witnesses are tangents to $\mathbb{F}_{\mathcal{M}}^\epsilon$. Minkowski's theorem \citep{int} states that every polytope in $\mathbb{R}^n$ is the convex hull of finitely many extreme points. Therefore, if there exists finitely many extreme points for a given convex and compact set, finitely many linear witnesses will suffice for separating all the entities outside that given set. The task is now to determine the number of extreme points of the set  $\mathbb{F}_{\mathcal{M}}^\epsilon$. We resolve this issue by proving the following theorem.

\textbf{Theorem 4:} The convex compact set of all MCS does not form a polytope.

\proof  Extreme points of a convex set are those entities within the set, having no non-trivial decompositions in terms of convex combinations of other points in the given set. Since the MCS are always valid physical states, the pure states, if there any, will lie on the vertices of the set. Hence, to prove the theorem, it is enough to show that there exists uncountable many pure Choi states. Consider the set of all Unitary channels: $\{\mathcal{U}_{\alpha}(t_2,t_1),~ \alpha \in \mathcal{I}\}$, where $\mathcal{I}$ is an uncountable index set. Unitary operations are divisible, i.e. $\mathcal{U}_{\alpha}(t_3,t_1)=\mathcal{U}_{\alpha}(t_3,t_2)\circ\mathcal{U}_{\alpha}(t_2,t_1)$, with $t_3\geq t_2\geq t_1$, $\forall~ t_1, t_2, t_3$. The Choi state corresponding to some Unitary operation $\mathcal{U}_{\alpha_0}(t_2, t_1)$, given by $\mathcal{C^{\mathcal{U_{\alpha_{0}}}}}=\mathbb{I}\otimes\mathcal{U_{\alpha_0}}(t+\epsilon, t)(\vert\phi\rangle\langle\phi\vert)$, is a pure maximally entangled state and hence it is an extreme point of $\mathbb{F}_{\mathcal{M}}^\epsilon$. Since $\mathcal{I}$ is uncountable, there exists uncountably many such pure maximally entangled states for given dimensions. Therefore there are uncountably many extreme points of $\mathbb{F}_{\mathcal{M}}^{\epsilon}$ and hence it does not form a polytope. \qed 

\begin{figure}[htb]
	{\centerline{\includegraphics[width=8cm, height=6cm] {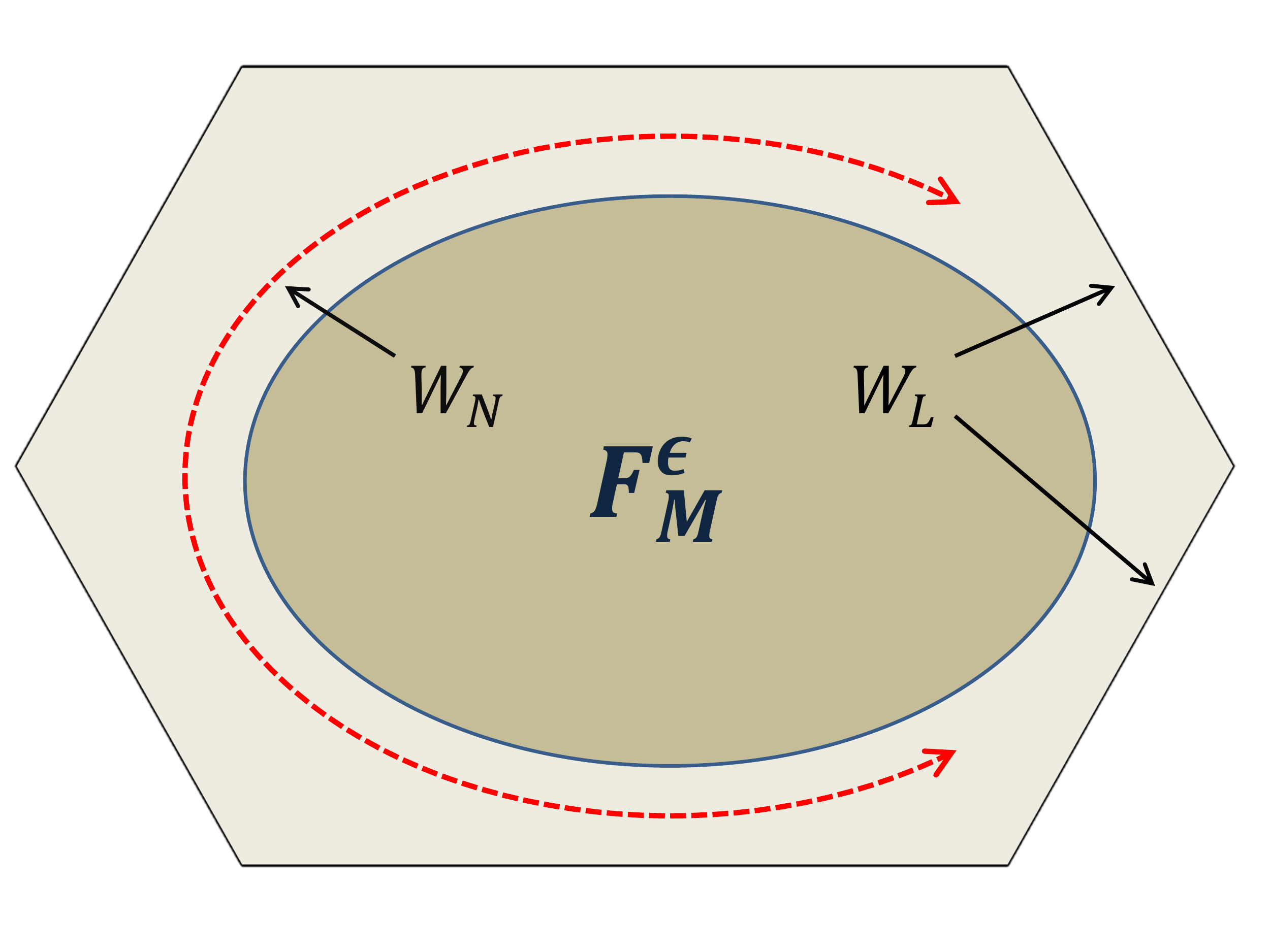}}}
	\caption{(Colour online) \\ The figure depicts a cross section of the set $\mathbb{F}_{\mathcal{M}}^{\epsilon}$. The outer polygon represents a convex set with a polytope structure, whose sides are optimal linear witnesses ($W_L$). Clearly we understand that if the set had a polytope structure, finite linear witnesses would have been sufficient to distinguish all the points not in the set. Since $\mathbb{F}_{\mathcal{M}}^{\epsilon}$ does not have that, as depicted by inner ellipse, nonlinear improvement is always possible. The red dashed line indicates such a non-linear witness $W_N$.   }
	\label{fig3}
\end{figure}

As a consequence of \textbf{Theorem 4}, we surmise the importance of nonlinear witnesses, to improve upon the efficiency of its linear counterpart. Since $\mathbb{F}_{\mathcal{M}}^{\epsilon}$ does not form a polytope, finite number of hyperplanes will not be sufficient to detect all the NMCS. Therefore nonlinear improvement of witnesses are necessary to detect them. In Fig. \eqref{fig3} we schematically justify the necessity of nonlinear witnesses.

From the above discussion, we have found that some of the pure maximally entangled states are in the set of all extreme points $Ext(\mathbb{F}_{\mathcal{M}}^{\epsilon})$ of the set $\mathbb{F}_{\mathcal{M}}^{\epsilon}$. We know that they are also among the extreme points of the state space $\mathbb{S}$. This fact tells us that $Ext\left(\mathbb{S}\right) \cap Ext(\mathbb{F}_{\mathcal{M}}^{\epsilon})$  is non-empty. But it is also evident that not all the pure maximally entangled states are in  $Ext(\mathbb{F}_{\mathcal{M}}^{\epsilon})$. This is because of the fact that the map is locally applied on one side of a bipartite maximally entangled state to construct the Choi states. It is therefore clear that the maximally entangled states generated by applying local unitaries on the other side, will not be among the set of Choi states. Those states, though among the extreme points of $\mathbb{S}$, will not be in $\mathbb{F}_{\mathcal{M}}^{\epsilon}$. The most obvious open question is then whether  $Ext(\mathbb{F}_{\mathcal{M}}^{\epsilon})$ is a strict subset of $Ext\left(\mathbb{S}\right) $. However, we make the conjecture that this is not the case. The argument behind this statement is the following. Since $\mathbb{F}_{\mathcal{M}}^{\epsilon}\subset\mathbb{S}$, there are valid physical states not contained in $\mathbb{F}_{\mathcal{M}}^{\epsilon}$. We have already proved that there exists pure states not contained in $Ext(\mathbb{F}_{\mathcal{M}}^{\epsilon})$. Therefore,  there can be mixed Choi states having no non-trivial state decomposition in terms of the pure states in $Ext(\mathbb{F}_{\mathcal{M}}^{\epsilon})$. Though they always have the same in terms of pure states which are in $Ext(\mathbb{S})$.

Let us consider the set of all Choi states $\mathbb{C}_{\mathcal{A}}$. It can be shown in a similar procedure of the proof of \textbf{Proposition 1}, that  $\mathbb{C}_{\mathcal{A}}$ is also a convex set under the small time interval approximation.  $\mathbb{F}_{\mathcal{M}}^{\epsilon}$ is of course a strict subset of  $\mathbb{C}_{\mathcal{A}}$. We have shown earlier that $\mathbb{F}_{\mathcal{M}}^{\epsilon}$ is also a strict subset of  $\mathbb{S}$. The set $\mathbb{F}_{\mathcal{N}}^{\epsilon}=\mathbb{C}_{\mathcal{A}} \setminus \mathbb{F}_{\mathcal{M}}^{\epsilon}$ contains all the NMCS. Clearly all the elements of  $\mathbb{F}_{\mathcal{N}}^{\epsilon}$ are not valid quantum states, because NM operations breaks CP-divisibility. Therefore, it is evident that $\mathbb{F}_{\mathcal{M}}^{\epsilon}=\mathbb{S}\cap\mathbb{C}_{\mathcal{A}}$. In the following Fig. \eqref{fig4}, we depict this discussion schematically.

\begin{figure}[htb]
	{\centerline{\includegraphics[width=8cm, height=6cm] {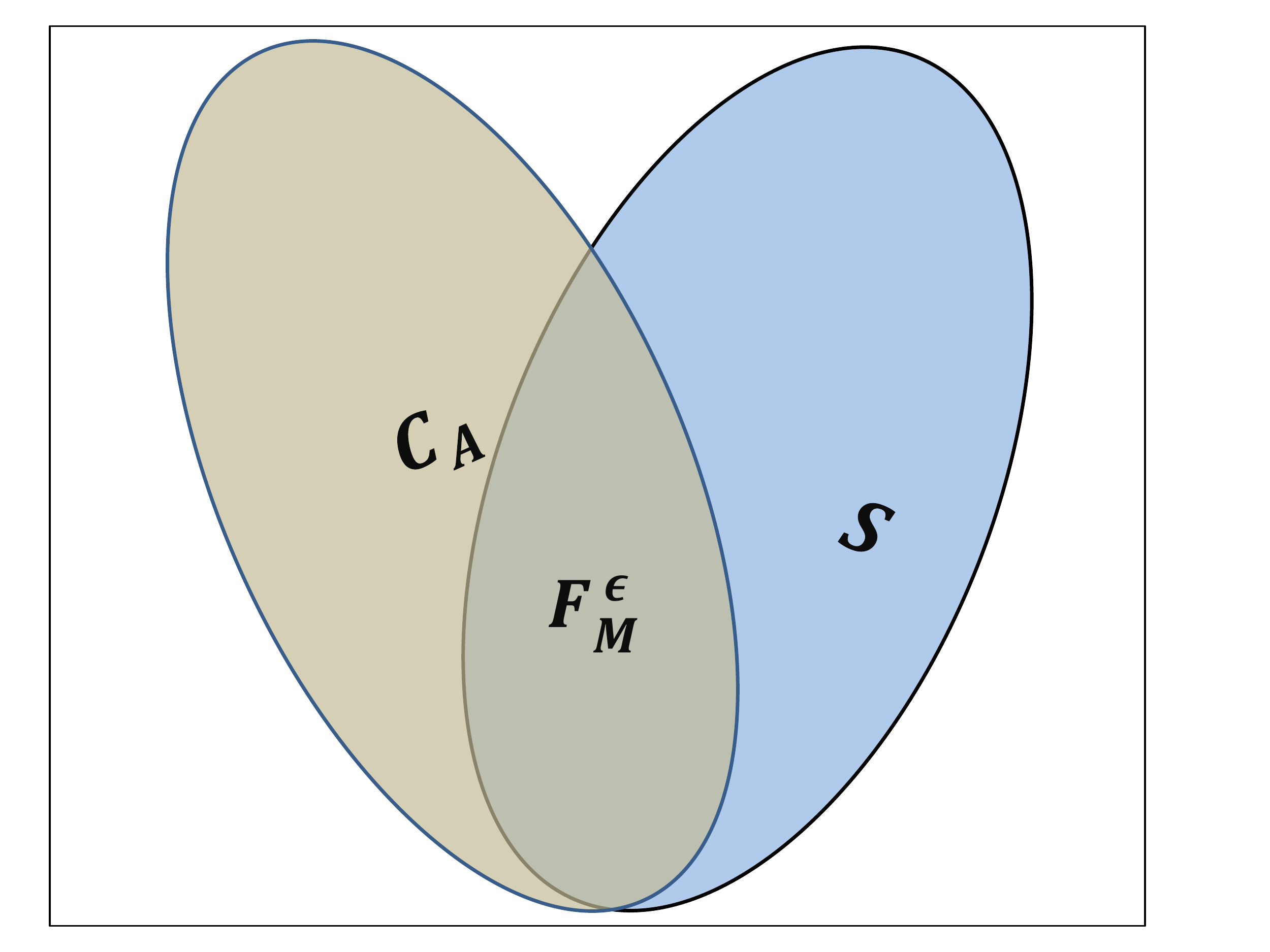}}}
	\caption{(Colour online) \\ Here we represent the convex sets $\mathbb{S}$ and $\mathbb{C}_{\mathcal{A}}$. Their intersection is another convex set  $\mathbb{F}_{\mathcal{M}}^{\epsilon}$. We can also understand from the diagram, that some of the extreme points of  $\mathbb{F}_{\mathcal{M}}^{\epsilon}$ are mixed density matrices. }
	\label{fig4}
\end{figure}

\textit{Conclusion:} In this letter, we develop a proper theory of NM witnesses, based on the convex and compact structure of the set of all MCS, under small time interval approximation. We construct an experimentally feasible framework of detecting non-Markovianity by hermitian witnesses, in snap-shots taken at different temporal regions for a given quantum evolution. We further investigate the geometric structure of the set of all MCS to find that they do not form a polytope, which opens up the possibility to consider nonlinear improvement of NM witnesses as a future line of investigation in the field of non-Markovian open system dynamics. 

\textbf{Acknowledgement:}
Authors thank Manik Banik of SNBNCBS, Kolkata for illuminating discussion. SB thanks SERB, DST, Government of India for financial support. BB thanks DST INSPIRE programme for financial support.

\bibliographystyle{apsrev4-1}
\bibliography{witness}

\end{document}